\documentstyle[aps,floats,epsf]{revtex}
\begin{document}
\setlength{\baselineskip}{4.2mm}
\draft
\preprint{STH-2001-01}
\title{
Universality in the periodicity manifestations in\\
the non-locally coupled map lattices in the turbulent regime\\} 
\author{Tokuzo Shimada, Shou Tsukada\\}
\address{Department of Physics, School of Science and Technology, Meiji University\\
Higashi-Mita 1-1-1, Tama, Kawasaki, Kanagawa 214-8571, Japan\\}
\maketitle
\begin{abstract}\\ \indent
In the turbulent regime of coupled map lattice with non-local interaction the maps systematically form 
periodic cluster attractors and their remnants by synchronization due to the foliation of periodic 
windows of the element map. 
We examine these periodicity manifestations in three types of coupled map lattices in $D=1, 2, 3$. 
In the first two, the interaction is all-to-all but the coupling decreases with distance in a power and an 
exponential law. In the third, the interaction is uniform but cut off sharply. 
We find that in all three models and in all dimensions periodicity manifests universally 
from turbulence when the same suppression of the local mean field fluctuation 
is achieved by the non-local averaging.\\
\end{abstract}

\noindent
{\bf 1. Introduction}\\ 

Synchronization of coupled chaotic elements was firstly shown for the chaotic flows 
in a master-slave relation \cite{pecora} and the more general forms of synchronizing chaotic flows have 
been extensively explored \cite{synchronization}. 
The synchronization occurs in general under a subtle balance 
between the randomness in the elements and the coherence forced by the interaction among them. 
In order to explore various phases of synchronization in complex systems of many chaotic elements, 
the coupled map lattice (CML) provides us with a concise testing ground. 
The CML with the nearest couplings exhibits the spatio-temporal chaos 
and pattern formation \cite{cml,kanekocml}.
 It's another limit, the globally coupled map lattice (GCML) with uniform all to all couplings, truncates 
the notion of the distance and features succinctly the battle between the order and the disorder \cite{kanekoGCML}.
 It is a basic model of the network of the chaotic neurons with clustering nature of synaptic connections, 
and also embodies the basic features of various physical systems 
such as coupled multi mode lasers, a Josephson junction array, fluid vortices and coupled electric circuits. 

The simplest GCML consists of $N$ identical chaotic maps and evolves in discrete time 
under an interaction via the mean field. It has only two parameters, the nonlinear parameter $a$ 
of the map and the coupling $\varepsilon$ of the averaging all-to-all interaction. 
Yet it is endowed with a rich variety of phases \cite{kanekoGCML}. 
If the coupling is taken very large, the maps synchronize in a single chaotic cluster.
In the intermediate coupling region, the maps form a few clusters by synchronization 
and suppress the fluctuation of their mean field. In the `turbulent regime'  --- the regime of very small 
coupling and high non-linearity --- no visible clusters are {\it in general} 
formed; for most of the coupling values the maps seem evolving randomly 
by direct observation. 
However, it has been recently found that, even at the very weak coupling in the 
turbulent regime, the maps systematically form periodic cluster attractors and their remnant
states by synchronization when a certain tuning condition between $a$ and $\varepsilon$ is  
satisfied \cite{ts,shibata,parravano,pre}. 
We call them {\it periodicity manifestations} (PM's) from the turbulence.

The PM's are intriguing synchronization phenomenon at very weak coupling. 
In the element map, many periodic windows are embedded in the chaos.  
The PM's are induced in GCML whose parameters are set  along the foliation curves 
of the periodic windows which run through the parameter space and there 
the dynamics of  maps is reduced by synchronization to that of a respective window.
It is known that the turbulent GCML with large $N$ is under unfailing weak coherence (so called hidden-coherence), which induces 
the violation of law of large numbers in the mean field fluctuation in time  
\cite{hiddencoherence}.
With the new findings of PM's, the turbulent GCML may be regarded as a system 
which sensitively mirrors the periodic windows of the element maps with the background weak coherence. 
The amazing fact that even at very weak coupling the maps easily form periodic cluster 
attractors may have important implications in complex systems of coupled chaotic elements, in 
particular in the activity of the brain. For instance, an efficient switch between the periodic states via chaos \cite{ogy} may be realized even in a system in the turbulent phase, provided that at least these PM's are 
not a particular phenomenon in the most simplified GCML. In this Letter we address ourselves to the question to what extent the PM's depend on the global coupling feature of it. 

Related works in the literature may be summarized as follows.
In one-dimensional CML it was shown that the hidden-coherence becomes visible 
with increasing coupling range \cite{sinha}; the phase diagram of the model at intermediate couplings 
was examined in \cite{kozma}, and the thermodynamical limit has been analytically investigated \cite{gade}.  
A two-dimensional CML with couplings following an inverse power law with the 
distance is shown to share the same phase diagram with GCML \cite{ts}. 
In CML of Ginzburg-Landau oscillators it is shown that the spatial correlator exhibits 
a power law decay \cite{kuramoto,nakao}.
It has been recently shown that in one-dimensional CML with power-law couplings 
the maximal Lyapunov exponent monotonically increases as the coupling-range varies from global to 
local \cite{batista}. 
These constitute sure progress in understanding the link  between the dynamics of GCML 
and CML. However, to date, the investigation has been mostly focused on the synchronous 
chaos and spatio-temporal pattern formation, rather than the formation of periodic clusters, and often limited to a particular dimension. 
In this Letter we turn our eyes to the PM's and investigate for the first time the variation of them
with the change of the coupling-range. We do this extensively in three non-local CML's in dimension $D=1,2,3$.  
All models interpolate the GCML and the nearest-neighbor CML but in different paths. 
We show that PM's occur in all models at sufficient non-locality.
Furthermore we report that there is a salient universality in PM's.  
They occur at the same strength, independent from detailed construction and the dimension
of the map lattices, when a factor ${\cal F}$, which represents the suppression of the local mean field 
fluctuation by averaging and calculated from the model parameters, is the same.%
\footnote{\footnotesize 
It is in form reminiscent of the universality in the Debye's theory of the specific heat, which
takes the same value over various crystals if compared at the same scaled temperature 
$T/\Theta_D$, where $\Theta_D$ may be evaluated from the phonon velocities in the crystals. }\\ \\

\noindent
{\bf 2. GCML,  the periodicity manifestations and the MSD curve} \\

The simplest GCML on the lattice $\Lambda$ is defined by an evolution equation
\begin{eqnarray}
   x_{P}(t+1)=(1-\varepsilon) f(x_{P}(t)) + \varepsilon h_t, ~P\in \Lambda,  
\label{gcmlevolution}
\end{eqnarray}
with the mean field $h_t \equiv \frac{1}{N} \sum_{Q \in \Lambda} f(x_{Q}(t))$ 
and $f(x)=1-a x^{2}$.
This is an iteration of a two step process; the independent mapping followed by an interaction 
via the mean field $h_t$ with an overall coupling $\varepsilon$.
By adding (\ref{gcmlevolution}) over $P$,
we find a relation $\frac{1}{N}\sum_{P \in \Lambda} x_{P}^\prime = h$;--- 
{\it the mean field is kept invariant in the interaction}.
All the non-local models below respect this invariance rule. 

The PM's are organized by the maps by synchronization when the parameters $a,\varepsilon$ 
are in a balance that allows a reduction of the high $N$-dimensional dynamics 
to that of the element logistic map in a periodic window at non-linearity $b$. 
The balance defines foliation curves on the $(a,\varepsilon)$ plane 
and all GCML on a curve are universally governed by the 
same window dynamics. 
The most prominent PM's are induced by the period three window \cite{ts}.  
If GCML is on the foliation curves of the $p3$ window, the maps organize themselves into 
almost equally populated three clusters, which oscillate mutually in period three --- $p3c3$ 
maximally symmetric cluster attractor (MSCA) \cite{pre}. 
With slightly higher $\varepsilon$ at the same $a$, that is, on the foliation curves 
from the intermittent region, the maps organize themselves in $p3c2$ cluster attractor. 
Both $p3$ attractors are formed at any reduction factor $(r \equiv b/a \le 1)$.
Generally, for a small reduction ($r \gtrsim 0.95$), maps form $p=c$ MSCA and $p>c$ clusters respectively 
along the curves of period $p$ window and in the nearby higher $\varepsilon$.
In MSCA, the MSD of the mean field fluctuation is minimized due to the high population symmetry 
and all observed MSCA's are linearly stable\cite{pre}.%
\footnote{\footnotesize
The stability is verified by algebraically solving the eigenvalue problem of the 
linear stability matrix and free from the numerical trap that occurs in the finite precision
iteration at the negative transverse Lyapunov exponent \cite{trap}. 
We have also verified that, for the non-local models in this note not at the 
GCML limit, the minimum gap among maps in a cluster reaches a plateau from above at $10^{-9}-10^{-10}$ and no trap occurs.}
Contrarily, in $p>c$ states, the MSD turns out extremely high due to missing clusters to fulfill the orbits.
For a large reduction ($r \lesssim 0.95$),  the clusters are no longer formed
but their remnants induce the same structure in the MSD ---
a valley and peak respectively along the curves of a window and in the nearby higher $\varepsilon$.
The sequence of the periodic windows in the element map therefore produces a 
successive valley-peak structure in the MSD curve (a function of $\varepsilon$ 
at $a$) and a given window induces a pair of a valley and a peak at the dictated position.%
\footnote{\footnotesize
The curve for a MSCA with a constant $h$ and produced by a window 
dynamics at $b$ is given 
$ (a , \varepsilon )^b(r) =\left (  \frac{b}{r}, 1 -   \frac{r y^*}{2}  -  \sqrt{   r (1- y^*) + \left(  \frac{ry^*}{2} \right )^2 } 
\right)$, 
where $y^*(b)$ is the time average of the single map orbit at non-linearity $b$.
}
We use below the peak-valley structure in the MSD curve as a succinct representation of PM's.
\\ \\ \\

%
\noindent
{\bf 3. Non-locally coupled map models}\\

\noindent
{\it 3.1. A power law model: POW$_\alpha$}\\

\begin{figure}[t]
\begin{center}
\leavevmode
\epsfxsize=73.33mm
\epsfysize=114.17mm
\epsfbox{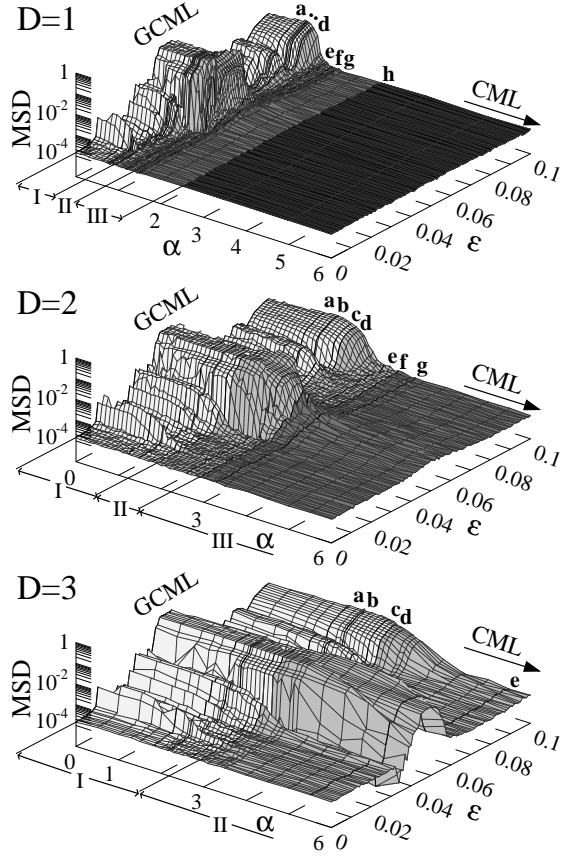}
\vspace{5mm}
{\footnotesize\caption{The MSD surfaces in POW$_\alpha$ over the $\alpha,\varepsilon$ plane. 
$a=1.90$ and $N=51^2$, $51^2$, $13^3$ for $D=1,2,3$ respectively.
The MSD curves indicated on the surfaces at the marking points 
are redisplayed in Fig.~2.
\label{msdsurfaces}}}
\end{center}
\end{figure}

\begin{figure}[t]
\begin{center}
\leavevmode
\epsfxsize=82.98mm
\epsfysize=71.4mm
\epsfbox{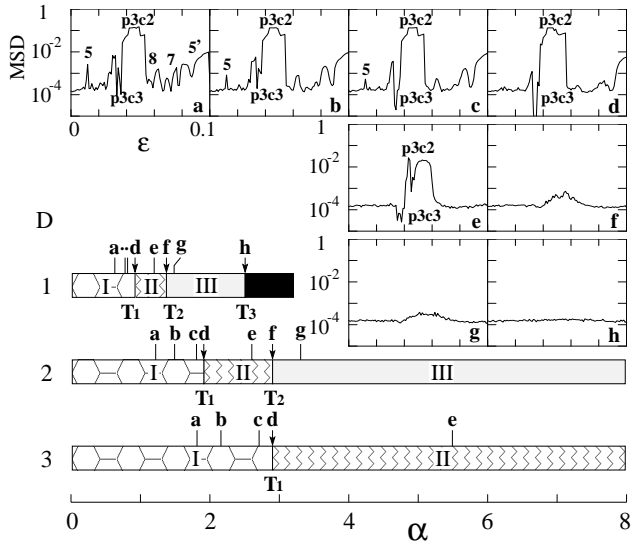}
\vspace{5mm}
{\footnotesize\caption{
The bar chart for the PM's in POW$_\alpha$ with $a=1.90$
and the MSD curves at the eight marking points (a-h).
I: full PM's, II: only the $p=3$ PM's,  III: only the hidden coherence. 
\label{barchartpow}}}
\end{center}
\end{figure}

As an extension of GCML let us consider a model  
\begin{eqnarray}
x_{P}^\prime 
&=& (1-\varepsilon) f(x_{P}) +\varepsilon h_P,  ~P \in \Lambda  \nonumber\\
h_P &\equiv& \sum_{Q \in \Lambda} W_{PQ} f(x_{Q}) \nonumber\\
&=&  c^{(\alpha)} f(x_{P}) +d^{( \alpha )} \sum_{\rho=1}^{\rho_{\max}} \frac{1}{\rho^{\alpha}} 
 \sum_{Q \in \Lambda_{\rho}(P)} f(x_{Q}), 
\label{genericevolution}
\end{eqnarray}
where each map couples to other maps via a {\it local mean field} $h_P$.
The $\Lambda_\rho (P)$ is a sub-lattice of $\Lambda$ consisting of maps at an equal distance 
$\rho$ from a site $P$.
For simple analytic estimates below, we approximate it 
by a set of points on the boundary of a $(2 \rho +1)^D$ square (cube) for $D=2(3)$.
The number of maps in $\Lambda_\rho$ is then given by  $n_\rho=2, 8 \rho, 24 \rho^2 +2$
for  $D=1, 2,3$ respectively.
We impose the periodic boundary condition and the maximum `radius' 
of $\Lambda_\rho$ is given by $\rho_{\max} =  (N^{1/D} - 1)/2$.
As a requisite the weights $W_{PQ}$ in $h_P$ must add to one; $\sum_{Q}W_{PQ}=1$. 
This, with the reciprocity $W_{PQ}=W_{QP}$,  leads to an important relation
$  \frac{1}{N}\sum_{P \in \Lambda} h_{P} = h$;---
{\it the average of the local mean fields is nothing but the mean field of the whole system},
which holds at any step of the iteration. This in turn guarantees the above invariance rule. 
From $\sum_{Q}W_{PQ}=1$,  it follows that 
\begin{eqnarray}
c^{(\alpha)} + d^{(\alpha)} S^{(\alpha)}=1, ~~S^{(\alpha)} \equiv \sum_{\rho=1}^{\rho_{\max}} \frac{n_\rho }{ \rho^\alpha}. 
\label{weightsumrule} 
\end{eqnarray} 
Let us make (\ref{genericevolution}) into a model which interpolates the GCML and the 
nearest neighbor CML.
In order to match with GCML at $\alpha=0$, the coefficient must be
$ c^{(0)}=d^{(0)}=1/N$.
In order to match with the nearest neighbor CML
\begin{eqnarray}
x^\prime_P = f(x_P)  +\frac{\varepsilon}{n_1+1} 
            \left( \sum_{Q \in \Lambda_1 (P) }f(x_{Q}) - n_1  f(x_P)   \right) 
\label{cml1}
\end{eqnarray}
 at $\alpha \rightarrow \infty$,  the coefficients must be $c^{(\infty)}=d^{(\infty)}={1}/({n_1+1})$.
In both limits, $c=d$. Therefore, 
we set $c^{(\alpha)}=d^{(\alpha)}$ for all $\alpha$ as the simplest interpolation. 
Normalizing the couplings by (\ref{weightsumrule}), 
we obtain a one parameter extension of GCML, POW$_\alpha$, with the local mean field
\begin{eqnarray}
 h_P=   \frac{1}{1+ S^{(\alpha)}} 
\left(
 f(x_{P}) 
	+ \sum_{\rho=1}^{\rho_{\max}} \frac{1}{\rho^\alpha} \sum_{Q \in \Lambda_{\rho}(P)} f(x_{Q}) 
\right). 
\label{POW}
\end{eqnarray}
We show in Fig.~\ref{msdsurfaces} and \ref{barchartpow} the result of our extensive analysis of 
the MSD of the time series of the mean field $h_t$ in POW$_\alpha$.
In Fig.~\ref{msdsurfaces} the MSD is shown as a surface plot over the $\alpha,\varepsilon$ plane. 
The surface is constructed for each $D$ by $100 (\varepsilon) \times 50 (\alpha)$ measured points 
over the time interval $t=10^3-2 \times 10^3$.  We have also taken the same amount of data for other two models below.
As noted in the previous footnote our analysis is immune from
numerical trap. The MSD curve at $\alpha=0$ is that of GCML and   
exhibits the foliation of various windows at full strength \cite{pre}. 
The PM's diminish with increasing $\alpha$ most quickly in $D=1$ and it is prolonged in higher dimensions.
Apart from this, the three surfaces are remarkably similar each other; 
if one picks a certain MSD curve at an $\alpha$ in, say, $D=1$,
one can also find the same curve in $D=2,3$ at some $\alpha^\prime$ and $\alpha^{\prime \prime}$. 
This was for us the first clue to the universality in PM's, which predicts the latter from $\alpha$. 

By the strength of PM's the $\alpha$ interval may be divided into three typical regions (I,II,III)
and for a closer analysis we have also chosen eight marking points (a-h),
which are indicated in both figures. In Fig.~\ref{barchartpow} we compare the phases in $D=1,2,3$ with respect to 
PM's by a bar-chart diagram and show the MSD curves at the marking points in the insets.

(I) One can observe all PM's that occur in GCML. 
From the GCML limit ($\alpha=0$) up to the point a, the full strength PM's are produced.  
The relevant dominant windows are marked on the MSD curve at a,
which agrees precisely with the curve in GCML \cite{pre}.
From a to d, the peak-valley structures, except for that due to $p3$
clusters, gradually diminish. The peak due to $p5$ window starts
diminishing at a and it becomes half-height at b. At c all the 
sub-dominant peak-valley structures vanish, and even the $p5$
peak vanishes at d. 

(II) The region of $p3$ PM's only. It starts from d
and the $p3c2$ peak disappears at e.  
At f, only a broad MSD peak is seen in the MSD curve. 

(III) Essentially the region of the hidden coherence. 
Only broad MSD peak can be seen around the foliation zone of the 
$p3$ window.  At the start of III and near the top of the peak,
the temporal correlator of maps decays in a $p3$ motion with exponential envelope.
At g, the broad MSD enhancement becomes half-height and 
the correlator fails to sense the periodicity everywhere. 
At h, the MSD enhancement disappears.

The transition points T$_1$, T$_2$, T$_3$ between 
the regions are d, f, h respectively. 
We note that $\alpha_{T_1} \approx 0.9,~1.9,~2.9$, 
approximately in the ratios $1:2:3$ for $D=1, 2, 3$ respectively. \\ \\

\noindent
{\it 3.2. A coupled map lattice with exponentially decaying couplings: EXP$_{\rho_0}$\\}

Similarly,  we obtain a model with the local mean field 
\begin{eqnarray}
&& h_P=\frac{1}{1+ S^{(\rho_0)}} \left( f(x_{P}) + \sum_{\rho=1}^{\rho_{\max}}w_{\rho, \rho_0} 
\sum_{Q \in \Lambda_{\rho}} f(x_{Q}) \right),  \nonumber \\
&& S^{(\rho_0)} \equiv  \sum_{\rho=1}^{\rho_{\max}} n_\rho w_{\rho, \rho_0}
\label{EXP} 
\end{eqnarray}
where $w_{\rho, \rho_0}\equiv \exp(-({\rho -1})/{\rho_0})$ is the exponentially decaying coupling.
This reduces to GCML at $\rho_0\rightarrow \infty$ and the nearest neighbor CML at $\rho_0 \rightarrow 0$.
The PM's diminish with {\it decreasing} $\rho_0$ via the same patterns of MSD curves.\\ \\

\noindent
{\it 3.3. A coupled map lattice with an interaction range $\kappa$: CML$_\kappa$}\\

\begin{figure}[t]
\begin{center}
\leavevmode
\epsfxsize=88.06mm
\epsfysize=62.9mm
\epsfbox{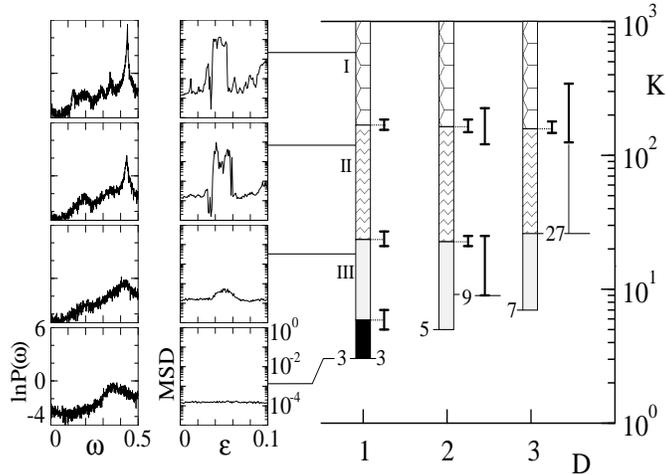}
\vspace{5mm}
{\footnotesize\caption{
Right: Bar chart of PM's in CML$_\kappa$ at $a=1.90$.  
I: full PM's, II: only the $p=3$ PM's, III: only the hidden coherence. 
The error bars compare the ambiguities in determining the transition points 
by two types of neighborhood.  
Center: The MSD curves at the pointed positions.
Left: Power spectrum of the mean field at $(a, \varepsilon)=(1.99,0.10)$ in $D=1$.
\label{barchartcmlk}}}
\end{center}
\end{figure}

Above two models maintain all-to-all coupling feature of GCML. 
Let us now consider a non-local CML with the local mean field
\begin{eqnarray}
h_P=\frac{1}{K} \left( f(x_P) + 
\sum_{\rho=1}^{\kappa}
\sum_{Q \in \Lambda_\rho (P) } f(x_{Q}) \right).  \label{cmlkappa}
\end{eqnarray}
Here $K=(2 \kappa+1)^D$ is the number of maps within a range $\kappa$.
We find that the PM's diminish with {\it decreasing} $\kappa$, again in the same process as above.  
Furthermore, we find that remarkably the same PM's occur irrespective to the dimensions if the 
neighborhood encloses the same number of maps. 
For instance,  the range $\kappa$ at T$_1$ is $77-92, 5-6, 2-3$
in $D=1,2,3$ respectively,  but  the number of maps $K$ 
within  $\kappa$ is $155-185$, $121-169$, $125-343$ in $D=1,2,3$.  
The large error in $D=3$ comes from the large-step increase of $K$ with $\kappa$. 
Thus for CML$_\kappa$ we determine the marking points by a refined neighborhood; 
a set of lattice points $Q$ around $P$ with $\sum_{i=1}^D (\Delta \rho_i)^2 \le \kappa^2$.
In Fig.~\ref{barchartcmlk} we show the three regions by a bar chart in terms of $K$. 
We find the bars in $D=1,2,3$ agree each other well. 

The $CML_\kappa$ in $D=1$ was used in an analysis of hidden coherence from the view of `beat' of 
mean field \cite{sinha}. The Fourier power spectrum of $h_t$ is reproduced also in Fig.~\ref{barchartcmlk} at 
$(a,\varepsilon)=(1.99, 0.10)$, which was chosen in \cite{sinha} to avoid a visible synchronization.
Interestingly,  the Fourier peaks due to the beat become outstanding in accord with the onset of PM's. 
The same holds at $(1.90,0.064 )$, which is an equivalent point via the foliation curve.%
\footnote{\footnotesize The synchronous chaos may disappear at the thermodynamical limit if the coupling range 
is fixed \cite{gade}. 
}
\\ \\

%

\noindent
{\bf 4. The universality of PM's in non-local models}\\

Let us now put above experimental results in an overview.

{\it 1. A working hypothesis. ---}
The difference between GCML and other non-local models is only in the interaction step.
In GCML, all the maps contract uniformly to $h$ by a factor $1-\varepsilon$, 
while in others a map $f(x_P)$ is contracted to the local mean field $h_{P}$
which distributes around the overall system mean field $h$. 
Therefore, when the variance of $\xi_P \equiv h_{P} -  h $ over the lattice is large in the evolution 
of the system, some distortion of map configuration must unavoidably be introduced in the interaction step. 
Contrarily, when the variance is small at each step of the iteration, 
such a distortion will be avoided and the non-local system may evolve just in the same way with GCML.  
Thus, it is natural to consider that the deviation from the global 
limit is controlled by the variance of $\xi_P$.

The $\xi_P$ is an weighted sum of maps of the form
\begin{eqnarray}
\xi_P=\sum_{Q \in \Lambda} \left( W_{PQ} -\frac{1}{N} \right) f(x_Q) \label{xip}.
\end{eqnarray}
where $W_{PQ}$ is the couplings in (\ref{POW}), (\ref{EXP}), (\ref{cmlkappa}).
If the spatial correlation between the maps are negligible, the variance of $\xi_P$
may be estimated at each time $t$ as
\begin{eqnarray}
 \langle \xi_P^2  \rangle_\Lambda  &\equiv& \langle (h_P-h)^2  \rangle_\Lambda
\approx  {\cal F} \langle (f_P - h )^2  \rangle_\Lambda
 \nonumber \\
{\cal F} &\equiv&  \sum_Q  (W_{PQ})^2 - \frac{1}{N} \label{calF},
\end{eqnarray}
where $\langle \cdots \rangle_\Lambda $ denotes the average over the lattice  
and $\sum_Q W_{PQ}=1$ is used.
The factor ${\cal F}$ represents the suppression of the variance of the $\xi_P$ 
by taking the weighted mean of map values over the lattice $\Lambda$.
At the global limit, $W_{PQ} \rightarrow 1/N$ and ${\cal F}\rightarrow 0$ (strictly no variance).
For intermediate couplings and large $N$ the factor $1/N$ may be neglected
and ${\cal F}$ is solely determined by the couplings. 
Combined with the above consideration let us propose a working hypothesis that 
{\it PM's occur universally in all non-local models when the factor ${\cal F}$ is the same} 
and put it under scrutiny below.

%
%
\begin{table}[b]
\caption{
The leading $N$ estimate of the ${\cal F}$ in POW$_\alpha$. \label{foliation}
}
\begin{tabular}{cccccccccc}
$D$ & $\alpha=0$& $\frac{1}{2}$ & $1$ & $\frac{3}{2}$ & 2  &$\frac{5}{2}$ & 3 & $\frac{7}{2}$ & $\infty$ \\
\tableline
\footnotesize 
1 &$0$&  $\ln N/4N$ & $\pi^2/12\ln^2N$  & $\zeta(3)/2\zeta^2(\frac{3}{2}) $ &
$ \cdots$ &$ \cdots$ &$ \cdots $&$ \cdots$ &$\frac{1}{3}$\\
2 & $0$&  $1/8N$ & $\ln N/4N$ & $\pi^2/96\sqrt{N}$ & $\zeta(3)/2 \ln^2 N$ &$\zeta(4)/8\zeta^2(\frac{3}{2})$ &
$ \cdots$ &$ \cdots $&$\frac{1}{9}$\\ 
3& $0$& $1/24N$&$1/3N$&$\ln N/4N$&$\pi^2/36N^{\frac{2}{3}}$&$\zeta(3)/48N^{\frac{1}{3}}$
& $\pi^4/240\ln^2N$&$ \zeta(5)/24\zeta^2(\frac{3}{2})$ &$\frac{1}{27}$
\end{tabular} 
\label{analyticsuppressionfactors}
\end{table}
\begin{figure}[t]
\begin{center}
\leavevmode
\epsfxsize=106.66mm
\epsfysize=83.14mm
\epsfbox{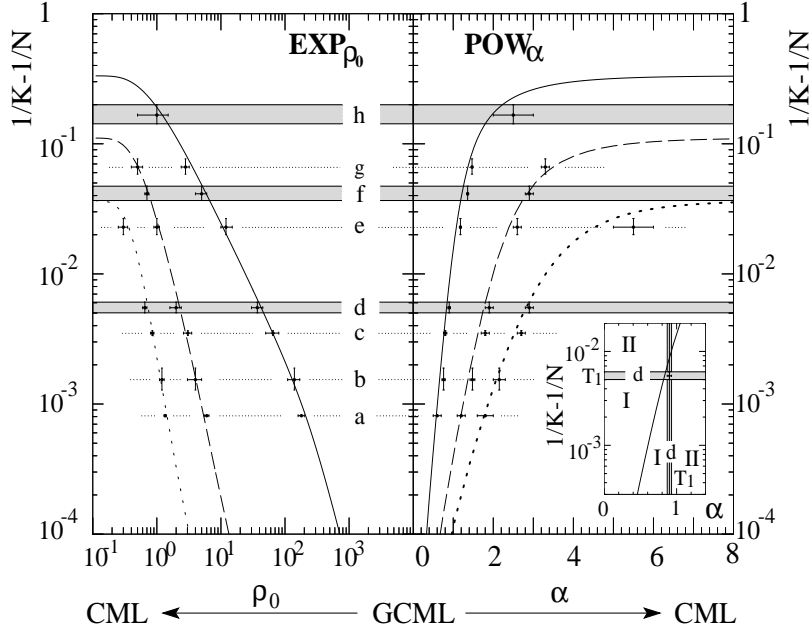}
\vspace{5mm}
{\footnotesize\caption{
The comparison between POW$_\alpha$ and CML$_\kappa$ (right)
and EXP$_{\rho_0}$ and CML$_\kappa$ (left) 
with respect to eight marked changes (a-h) of the PM's.
For $D=1,2,3$ respectively the ${\cal F}_D$ is shown by 
the solid, dashed, and dotted curves and $N=51^2, 51^2, 13^3$.
The inset illustrates the method for the case of d (T$_1$) in $D=1$.
The $\alpha$ ($\rho_0$) axis is in the normal (logarithmic) scale.
\label{modelcomparison}}}
\end{center}
\end{figure}

{\it 2. The factor ${\cal F}$ in each model.---} 
In CML$_\kappa$, the suppression factor ${\cal F}$ is simply ${\cal F}=1/K -1/N$.
This succinctly explains the observation that PM's occur with the same strength at common $K$ 
in all $D$ and uniformly diminish with decreasing (increasing) $K$ (${\cal F}$).

In POW$_\alpha$ the ${\cal F}$  is given by  
\begin{eqnarray}
 {\cal F}_D^{(\alpha)}=\frac{1}{(1+S^{(\alpha)}_D)^2}\left(1+\sum_{\rho=1}^{\rho_{\max}}             
  \frac{n_{\rho,D}}{\rho^{2\alpha}}\right) - \frac{1}{N}. 
\label{POWsuppfac}
\end{eqnarray}
The leading $N$ estimates for ${\cal F}_D^{(\alpha)}$ are tabulated in Table.~\ref{analyticsuppressionfactors}. 
We find in particular  ${\cal F} \approx \log N/4N$ at $\alpha=1/2,1,3/2$ for $D=1,2,3$ respectively.  
This gives a prediction that the PM's would be universal among 
POW$_{\alpha=1/2}^{D=1}$, POW$_{\alpha=1}^{D=2}$ and POW$_{\alpha=3/2}^{D=3}$. 

This is indeed the case; the full strength PM's, the same with those in GCML, are realized in all the three. 
The ${\cal F}_D^{(\rho_0)}$ in EXP$_{\rho_0}$ may be obtained by substituting 
$S_D^{(\rho_0)}$ and $w^2_{\rho,\rho_0}$ to $S_D^{(\alpha)}$ and $1/\rho^{2\alpha}$ respectively. 

{\it 3. Overall comparison of the models.---}
In Fig.~\ref{modelcomparison} we compare POW$_\alpha$ (EXP$_{\rho_0}$) with CML$_\kappa$
 in the right (left). The inset illustrates the case of the transition point T$_1$(d) in $D=1$
as an example.  The curve is ${\cal F}$ for POW$_\alpha$ in (\ref{POWsuppfac}).
The measured $\alpha$ at T$_1$ gives the vertical band taking account for the ambiguity 
in judging the MSD curve pattern.  
Hence,  the crossing of the curve and the vertical band gives the estimate of ${\cal F}$ in POW$_\alpha$ 
at its T$_1$ in $D=1$. 
On the other hand, ${\cal F}$ is universal over $D$ in CML$_\kappa$.
The  ${\cal F}$ at T$_1$ of CML$_\kappa$ gives the horizontal band, again 
counting for the ambiguity and averaged over $D$.
Thus, the vertical axis is used for both ${\cal F}$'s, that in POW$_\alpha$ and that in CML$_\kappa$.
If both models share exactly the same ${\cal F}$ at T$_1$, the curve will pass through the crossing junction
of the horizontal and vertical bands. In this example, the curve crosses 
the vertical band slightly above the junction
and the estimated ${\cal F}$ are $(8.5 \pm 1.5) \times 10^{-3}$ 
and $(5.5 \pm 0.5) \times 10^{-3}$ 
in POW$_\alpha$ and CML$_\kappa$ 
respectively.  Or, one can predict the $\alpha$ at T$_1$ in 
POW$_\alpha$ from $K$ at T$_1$ in CML$_\kappa$ using
the ${\cal F}$ curve of POW$_\alpha$. 
The prediction is $0.83 \pm 0.02$ to be compared with the measured $0.90 \pm 0.03$. 
In the overall comparison, only the junctions are shown by error-bars. 
We find that the hypothesis remarkably works with respect to all the marking points and in $D=1,2,3$
for ${\cal F}$ ranging from $10^{-4}$ to $O(1)$.

A few remarks are in order. \\
{\it (a) The 1:2:3 rule in POW$_\alpha$;---}
The curves in POW$_\alpha$ agree approximately with each other 
after a scale transformation $1:1/2:1/3$ in $\alpha$ up to ${\cal F}_D \approx 2 \times 10^{-2}$.
This extends the rule obtained by the leading $N$ calculation. 
The same strength PM's occur up to the marking point e if $\alpha$ is 
in the ratio of the system dimensionality, just like the universal PM's at the same $K$ in CML$_\kappa$.  \\
{\it (b) Missing transition points;---}
The horizontal bands exhibit three transition points observed in 
CML$_\kappa$ using refined neighborhood. 
In other models with the coarse neighborhood, the T$_3$ is missing in $D=2$ and 
both T$_2$ and T$_3$ are missing in $D=3$. (See Fig.~\ref{barchartpow}
for POW$_\alpha$). 
The curves of ${\cal F}$ explain the difference succinctly; 
they are constrained by the limiting values $1/3^D$ so they can pass through 
only the lowest two (one) bands in $D=2(3)$. We have numerically checked 
that the missing transition points are retrieved in both POW$_\alpha$ and EXP$_{\rho_0}$
with the refined neighbors.
\begin{figure}[t]
\begin{center}
\leavevmode
\epsfxsize=82.17mm
\epsfysize=75.51mm
\epsfbox{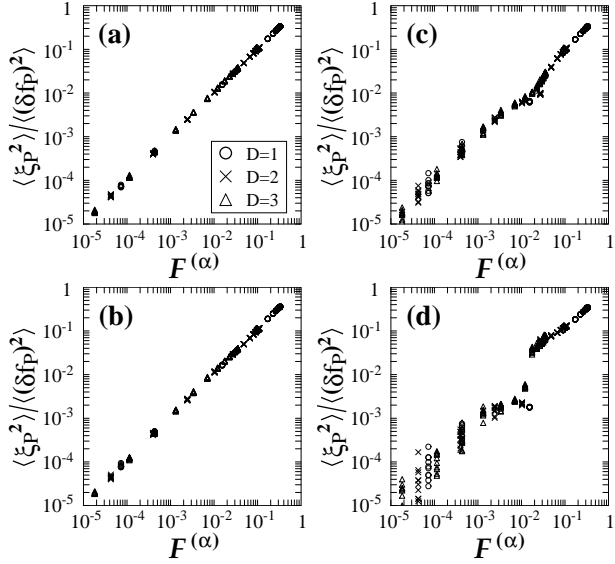}
\vspace{5mm}
{\footnotesize\caption{
The ratio $\langle (\xi_P)^2  \rangle_\Lambda / \langle (f_P - h )^2  \rangle_\Lambda$
averaged over 100 steps versus ${\cal F}^{(\alpha)}$, for $\alpha =0.5-8.0$ 
with $\Delta \alpha=0.5$ (increasing ${\cal F}^{(\alpha)}$) in POW$_\alpha$.
(For $D=1$, $\alpha=0.3$  is added.) Points from 10 random starts are overlayed.
 (a),(b),(c),(d) for $\varepsilon$= 0.02, 0.08, 0.0352, 0.045 respectively.
\label{Fcheck}}}
\end{center}
\end{figure}

\noindent
{\it (c) Finite-N effect;---}
Let us discuss the case of POW$_\alpha$.
As the leading $N$ estimate in Table~\ref{analyticsuppressionfactors} shows, 
${\cal F}$ vanishes at the thermodynamical limit
if $\alpha \le 1, 2, 3$ in $D=1,2,3$ respectively, while 
for the larger $\alpha$ it approaches a constant.  
Now, we have already chosen $N >10^3$, since it is necessary for the PM's 
(except for $p3$) to occur even in GCML. Thus the marking points a and b 
are already deep in region I and insensitive to a further increase of $N$. 
The points e to h are also insensitive because of the asymptotic limit of ${\cal F}$ at a constant. 
Hence the only place to check the finite size effect is the region near the T$_1$
and we have verified it (${\cal F} \propto 1/\log^2N$). 
For instance, T$_1$ in $D=2$ POW$_\alpha$ 
with $N=51^2$ is at $\alpha \approx 1.9$ and by replacing $N$ by $10^5$
the PM's become into stronger ones,
the type around point c in region I, with respect to the $N=10^5$ GCML
\footnote{\footnotesize
In GCML, the peak-valley structure becomes out-standing with increasing $N$
and remains the same for $N=10^5-10^6$\cite{pre}.
},
just as predicted.    
 \\
{\it (d) Approximation check;---} 
Our estimate of ${\cal F}$ is based on an approximation that the spatial correlation is negligible.
We have checked that this is legitimate. Firstly, we note that the even at the formation of 
cluster attractors such as $p3c3$ MSCA and $p3c2$ states, the spatial distribution of maps 
does not show any visible clustering. To avoid a confusion we stress that the clustering is in the map values 
and not in the spatial distribution.  We have checked that over 
the whole turbulent regime of the three models no visible spatial clusters are formed. 
Furthermore we have directly checked that the estimation in (\ref{calF}) works 
remarkably well in three models.%
\footnote{\footnotesize
Recently an interesting anomalous power-law spatial correlation in the Ginzburg-Landau 
oscillators is found in \cite{kuramoto,nakao}. For $\varepsilon=0-0.1$ at $a=1.90$, the range for GCML turbulent regime, 
the nearest neighbor CML is in fully chaotic phase \cite{kanekocml}.
But for the non-local CML's, we consider a similar analysis is 
necessary to confirm our approximation.}
The case for POW$_\alpha$ is presented in Fig.~\ref{Fcheck}.   
The left two boxes are the results for $\varepsilon=0.02, 0.08$,
where no visible synchronization occurs, and the right two for $\varepsilon=0.0352, 0.045$
where the $p3c3$ and $p3c2$ states are formed respectively.
We find that in all dimensions and for all coupling ranges the estimate works remarkably well
for ${\cal F}$ from $10^{-5}$ to $O(1)$, which fully covers the range of ${\cal F}$ in Fig.~\ref{modelcomparison}.
The spread of the data points observed in $p3c2$ cluster attractor  
is due to the different ratios of map populations in the two clusters 
formed from different initial configurations.\\
{\it (e) Inhomogeneous map lattices;---} 
Even if inhomogeneity is introduced randomly to the non-linearity of maps 
($a_P \rightarrow a_P \pm \delta a_P$), all features of PM's discussed in this note 
are unchanged for $\delta a_P < \delta a=0.01$.
For larger $\delta a$ the effect is similar to that due to  the decrease of non-locality.
For instance, with $\delta a=0.03$, the sequence of marking points starts from 
the point d for $D=1,2,3$.  We have also tested that the PM's occur in non-local CML's with other types of maps with successive windows. 
Details will be discussed elsewhere. \\ \\

\noindent
{\bf 5. Conclusion}\\

In this note we have focused our attention to the recently found periodicity manifestations in the 
turbulent regime of GCML. We have conducted an extensive statistical analysis 
in three non-locally coupled map lattices over $D=1,2,3$  
and examined to what extent they depend on the non-locality of the models.   
We have noted that the essential deviation of the non-local CML
from the GCML stems in the variance of the local mean field around the overall mean field.
We have analytically estimated the suppression factor ${\cal F}$ of the variance 
under an approximation that the spatial correlation of maps in the turbulence 
regime is negligible and checked that this ${\cal F}$ remarkably 
agrees with the numerical result. 
We have found a salient universally that, irrespective of the difference in construction and
the dimension of the lattice, the periodicity manifestations occur at the same strength to a good approximation 
once ${\cal F}$ is the same.
\\ \\

\noindent
{\bf Acknowledgements}\\

One of authors (T.S.) especially thanks Wolfgang Ochs for encouragement and reading the manuscript, and Max-Planck Institut f\"ur Physik, M\"unchen for the warm hospitality during his visit.  Our thanks also go to Mario Cosenza for communication and Kengo Kikuchi for collaborating with us at the early stage of this work. 


\end{document}